# A lightweight deep learning pipeline with DRDA-Net and MobileNet for breast cancer classification

Mahdie Ahmadi[1], Nader Karimi[1], Shadrokh Samavi[2]

[1]Dept. of Electrical and Computer Engn, Isfahan University of Technology, Iran
[2]Dept. of Computer Science, Seattle University, USA

*Abstract*— Accurate and early detection of breast cancer is essential for successful treatment. This paper introduces a novel deep-learning approach for improved breast cancer classification in histopathological images, a crucial step in diagnosis. Our method hinges on the Dense Residual Dual-Shuffle Attention Network (DRDA-Net), inspired by ShuffleNet's efficient architecture. DRDA-Net achieves exceptional accuracy across various magnification levels on the BreaKHis dataset, a breast cancer histopathology analysis benchmark. However, for real-world deployment, computational efficiency is paramount. We integrate a pre-trained MobileNet model renowned for its lightweight design to address computational. MobileNet ensures fast execution even on devices with limited resources without sacrificing performance. This combined approach offers a promising solution for accurate breast cancer diagnosis, paving the way for faster and more accessible screening procedures.

## 1. Introduction

As the fight against breast cancer continues, early and accurate diagnosis remains paramount. Medical professionals increasingly use advanced imaging techniques to gain valuable insights into the disease. At the same time, ultrasound scans and X-rays offer a window into the body's internal structures, and histopathological images (HIs) reign supreme [1]. These detailed microscopic views of tissue samples, obtained through biopsies, provide pathologists with unparalleled detail regarding cell morphology and organization. This information is crucial for forming accurate diagnoses and guiding treatment decisions. However, the traditional manual analysis of HIs is a labor-intensive process. It requires highly trained pathologists, often in limited numbers, to spend significant time meticulously examining slides under a microscope. This approach can be prone to human error and inconsistencies, particularly workloads and fatigue. This research classifies histopathology images into two classes: benign and malignant. For example, Fig 1 presents two images.

Numerous noteworthy approaches have been introduced in breast cancer histopathological image classification. In some research studies, transfer learning is employed for image classification tasks in medical imaging due to the challenges of acquiring large medical datasets containing sufficient and necessary features, often not readily available to the public. Transfer learning generally yields better performance and greater flexibility as it leverages a diverse and extensive set of images for training [2].

In the method proposed by Fan et al. [2], a Support Vector Machine (SVM) is utilized for classification, and the AlexNet architecture, pre-trained on the ImageNet dataset containing histopathology images, is employed. Similarly, Ferreira et al. [3] use the Inception-ResNet-v2 architecture, pre-trained on ImageNet, as the core network. They remove the initial layers of the pre-trained network to avoid overfitting specific training data features, as these layers are highly dependent on the training process and data. Finally, softmax is used as the classifier in their approach. Deniz et al. [4] utilize the AlexNet architecture for feature extraction from images in the BreakHis dataset. Instead of removing the initial layers, they removed the last three layers and added new layers, training the modified network and employing SVM for classification. In the method proposed by Vesal et al. [5], two pre-trained networks, ResNet50 and Inception v3, trained on the ImageNet dataset, are utilized. These networks are fine-tuned on the BACH2018 challenge dataset. ResNet50 is selected for its residual learning framework, while Inception v3 is chosen for its factorized inception modules. Also, the approach described by Mahesh Gour et al. [1] employs fine-tuned VGG16 and VGG19 models with 5-fold cross-validation and image augmentation.

These approaches highlight the diverse strategies in utilizing transfer learning for medical image classification, leveraging pre-trained architectures and fine-tuning them on specific medical datasets to achieve improved performance.

There are AI-based and deep learning methods that do not rely on pre-trained networks. Instead, they innovate by creating new networks specifically for classifying histopathology images. Yu et al. [6] introduced CA-BreastNet, an improved DenseNet neural network augmented with a coordinated attention mechanism. In the approach presented by Singh et al. [7], they utilize residual blocks inspired by the ResNet architecture and inception blocks inspired by the Inception-v3 architecture. The strength of this network enables it to achieve promising results on both the large dataset BHI and the small dataset BreaKHis [8]. Erfankhah et al. [9] proposed heterogeneity-aware multi-resolution LBP (hmLBP) leverages rotation-invariant uniform LBP as a foundation, utilizing multi-resolution analysis to effectively capture texture patterns in histopathology scans and address the challenge of polymorphism in diagnostic pathology. In the method proposed by Hu et al. [10], they employ stain normalization and data augmentation to enhance the performance of their innovative network. They refine a convolutional neural network using residual learning in their approach, which leads to improved results. These methodologies offer valuable insights into the diverse breast cancer histopathological image classification strategies.

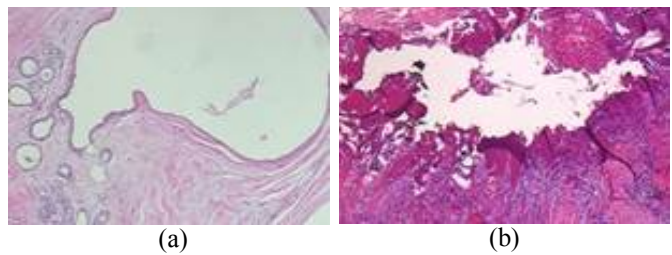

(a)           (b)

Figure 1: Hematoxylin and lesion-stained breast histopathology images (a) benign lesion, (b) malignant lesion.

In this study, amidst the diversity of deep learning models, convolutional neural networks (CNNs), particularly DenseNet, have emerged as prominent tools for automating the analysis of medical images in breast cancer diagnosis. The existence of a channel attention mechanism aids in learning complex patterns, yet overfitting and underfitting issues persist. These challenges are addressed by the densely connected blocks within the model.

In the proposed method, the devised network utilizes the MobileNet architecture, pre-trained on the ImageNet dataset, to extract features from images. Subsequently, this pre-trained MobileNet model is fine-tuned on the BreakHis dataset. The extracted features are then passed to the next part of the network, inspired by the DRDA-Net [11] architecture, for analysis and decision-making. The DRDA-Net network leverages the ShuffleNet architecture to enhance results and learn complex patterns in histopathology images. Our approach improves classification accuracy and operational speed by leveraging pre-trained networks for feature extraction and streamlining the model's complexity. Through detailed insights into the model's design, dataset description, experimental challenges, and rigorous evaluation of results, this study contributes to advancing histopathological image analysis for breast cancer diagnosis.

The structure of this paper is as follows: Section 2 outlines our methodological approach and provides a comprehensive breakdown of the model's design, including its layers, components, and the rationale behind architectural choices. In section 3 ,we examine the results obtained from the proposed method and compare them with other approaches based on various criteria. The conclusion is presented in section 4.

## 2. Proposed Method

In contemporary machine learning, CNNs are pivotal for image classification, yet their increasing complexity poses challenges in efficiency, computational cost, and interpretability. Dense connections in deep CNNs exacerbate parameter explosion and overfitting. Recent innovations like group and depth-wise convolution address parameter efficiency but suffer from inter-group feature communication limitations. Shuffle networks mitigate this bottleneck, notably ShuffleNet, balancing efficiency and accuracy. Our backbone model, DRDA-Net, integrates ShuffleNet for optimized parameterization and feature exchange. Additionally, we adopt feature extraction from pre-trained networks MobileNet to reduce model dimensions training costs and enhance accuracy. Further details will be elaborated in subsequent sections.

This section introduces the architecture designed for classifying breast histopathology images into benign and malignant categories. Our approach utilizes the DRDA-Net as the backbone, augmented with MobileNet for initial feature extraction. The entire workflow of the proposed method is shown in Fig. 2. The DRDA-Net offers superior performance compared to other models, notably due to its parallel processing capability, leading to faster processing speeds crucial for efficient analysis of histopathology images. MobileNet, chosen for its lightweight design and efficient resource utilization, efficiently extracts discriminative features from input images, a critical step for subsequent classification tasks.

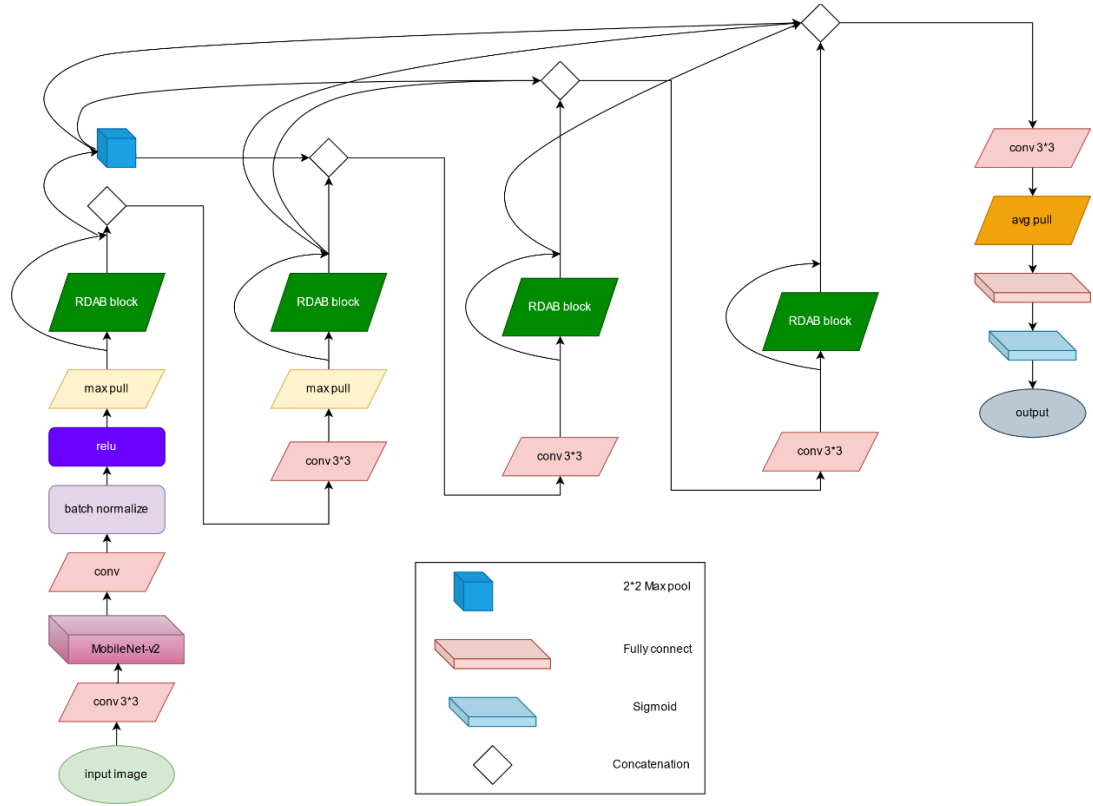

Figure 2: Block diagram of the proposed model

Following the feature extraction phase, the extracted features are forwarded to subsequent layers, consisting of convolutional and max-pooling layers. These layers aim to refine the feature representations further, capturing spatial hierarchies and enhancing discriminative characteristics within the data. Normalization is applied to the output of these convolutional and max-pooling layers to ensure stability and effective propagation of features through the network. This normalization step is pivotal in maintaining consistent gradients during the training process, thereby facilitating convergence and improving the overall robustness of the model.

Subsequently, the processed features are channeled through a series of residual dual-shuffle attention blocks (RDAB). Derived from the RDAB block architecture Fig. 3. these blocks incorporate residual connections. Residual connections are pivotal in mitigating the vanishing gradient problem, a common challenge in training deep neural networks. However, their introduction may introduce heightened model complexity as additional connections and parameters are integrated, potentially impacting computational efficiency and memory requirements. Moreover, while residual connections are instrumental in enhancing gradient flow and facilitating training, they may concurrently engender the risk of overfitting, particularly in scenarios where model capacity surpasses the complexity of the dataset. Hence, it becomes imperative to employ meticulous regularization techniques to counterbalance this propensity for overfitting.

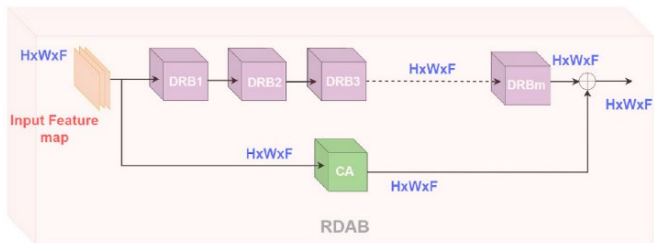

Figure 3: Architecture of the residual dual-shuffle attention block [11]

Following the RDAB blocks, additional convolutional and avg-pooling layers are employed to refine the feature representations further, extracting higher-level abstractions from the input data. This is followed by a fully connected layer, which integrates the extracted features into a compact representation suitable for classification. Finally, a sigmoid activation function facilitates binary classification, distinguishing benign and malignant cases based on the learned feature representations.

Due to the utilization of MobileNet as a feature extractor, there is no need for max-pooling layers to extract feature sizes. Therefore, most of the max-pooling layers are removed from the base network to preserve the overall features of each image, enabling better decision-making processes.

## 3. Experimental Results

*A. Image dataset*

The proposed model in this study aims to classify breast cancer using histopathology images, utilizing the BreaKHis dataset. The BreakHis dataset is a crucial resource in breast cancer research, comprising 7,909 histopathological images categorized into benign and malignant tumors across four magnifications(40x, 100x, 200x, 400x), as shown in Table. 1. With images meticulously sized at 752 × 582 pixels, it offers a diverse collection spanning eight cancer types. Despite its significance, the dataset's modest size poses challenges, with each magnification level containing approximately 2,000 images.

*B. Evaluation*

This diversity aids in robustness assessment. Performance analysis across these magnification levels was conducted, evaluating accuracy (1), Precision (2), Recall (3), F1 score (4), and test time to assess accuracy and inference time comprehensively. We show the test result in Table. 2.

$$Accuracy = \frac{100 * number\ of\ correct\ labels}{number\ of\ total\ labels} \quad (1)$$

$$Precision = \frac{true\ positive}{(true\ positive + false\ positive)} \quad (2)$$

$$Recal = \frac{true\ positive}{(true\ positive + false\ negative)} \quad (3)$$

$$F1\ score = \frac{2 * precision * recall}{precision + recall} \quad (4)$$

Table 1: BreaKHis dataset data distribution

| Magnification | Benign | Malignant | Total |
|---|---|---|---|
| 40x | 598 | 1398 | 1996 |
| 100x | 642 | 1437 | 2079 |
| 200x | 594 | 1418 | 2012 |
| 400 x | 590 | 1232 | 1822 |
| total | 2424 | 5485 | 7909 |

The experimentation pipeline is implemented using the PyTorch library. The input images are zero-padded to the 212 × 212 for training the model. It takes around 30 minutes for 30 epochs with an Nvidia GeForce RTX 3050 GPU with batch size 32. The time spent on this process by the DRDENET model is 2.2 times more than the time spent by our method. Experiments on validation datasets across magnification levels were conducted to determine the optimal value of the parameter "m" representing the number of DRA blocks within RDBA. Findings indicated superior performance when m was set to 1 compared to other values, resulting in a streamlined model architecture with decreased complexity. This optimization expedited training, reduced computation per iteration, and enhanced model interpretability and generalization capabilities, underscoring the importance of parameter selection and architectural refinement in optimizing model performance for breast cancer classification tasks.

Table 2: Test results for different magnification levels

| Metrics | 40× | 100× | 200× | 400× |
|---|---|---|---|---|
| Accuracy % | 97.92 | 97.03 | 97.03 | 97.90 |
| Precision % | 99.48 | 99.32 | 99.22 | 98.32 |
| Recall % | 97.46 | 96.31 | 96.46 | 98.59 |
| F1 % | 98.46 | 97.79 | 97.82 | 98.46 |
| Test time (ms) | 32.62 | 30.05 | 32.10 | 26.75 |

Furthermore, we compared our results with large and heavy pre-trained networks and newly developed convolutional networks that have worked in this area. It is observed that our network has performed better in all metrics. You can refer to Table 3 for the comparison results.

TABLE 3: PERFORMANCE COMPARISON WITH OTHER MODELS. ACCURACY, PRECISION, RECALL, AND F1 COMPARE DIFFERENT METHODS.

| Model | Metric | 40× | 100× | 200× | 400× |
|---|---|---|---|---|---|
| VGG19 | Acc | 92.11 | 92.00 | 92.00 | 93.45 |
|  | Pre | 93.21 | 91.15 | 90.96 | 94.00 |
|  | Recall | 91.65 | 93.44 | 91.11 | 91.17 |
|  | F1 | 92.42 | 92.88 | 91.03 | 92.56 |
| ShuffleNet | Acc | 93.10 | 89.47 | 95.54 | 90.41 |
|  | Pre | 91.95 | 91.77 | 94.31 | 87.89 |
|  | Recall | 92.00 | 90.56 | 93.44 | 89.54 |
|  | F1 | 91.97 | 91.16 | 93.87 | 88.71 |
| ResNet | Acc | 94.97 | 93.33 | 94.10 | 92.79 |
|  | Pre | 94.44 | 92.11 | 92.31 | 91.78 |
|  | Recall | 93.00 | 93.00 | 91.45 | 90.25 |
|  | F1 | 93.71 | 92.55 | 91.88 | 91.01 |
| DenseNet169 | Acc | 92.17 | 91.19 | 92.44 | 91.99 |
|  | Pre | 92.00 | 90.21 | 91.31 | 89.89 |
|  | Recall | 91.00 | 88.44 | 92.06 | 87.41 |
|  | F1 | 81.12 | 89.32 | 91.68 | 88.63 |
| Gour et al [1] | Acc | 82.12 | 82.98 | 80.85 | 81.83 |
|  | Pre | 95.07 | 91.59 | 85.25 | 88.59 |
|  | Recall | 86.39 | 86.98 | 90.80 | 88.53 |
|  | F1 | 90.49 | 89.20 | 87.57 | 88.38 |
| DRDA-net [11] | Acc | 95.72 | 94.41 | 97.43 | 96.84 |
|  | Pre | 94.00 | 96.00 | 96.00 | 98.10 |
|  | Recall | 96.90 | 93.20 | 99.00 | 95.20 |
|  | F1 | 95.40 | 94.60 | 97.44 | 96.62 |
| Erfankhah et al [9] | Acc | 88.3 | 88.3 | 87.1 | 83.4 |
|  | Pre | - | - | - | - |
|  | Recall | - | - | - | -- |
|  | F1 | - | - | - | - |
| Our method | Acc | 97.92 | 97.03 | 97.03 | 97.90 |
|  | Pre | 99.48 | 99.32 | 99.22 | 98.32 |
|  | Recall | 97.46 | 96.31 | 96.46 | 98.59 |
|  | F1 | 98.46 | 97.79 | 97.82 | 98.46 |

## 4. Conclusion

In this research work, we achieved significantly better results on the BreaKHis dataset through a two-pronged strategy prioritizing accuracy and efficiency. Firstly, we implemented a pre-trained MobileNet model. MobileNet's architecture boasts a relatively low number [3] of parameters, making it suitable for execution on devices with limited computational resources. This is particularly advantageous for real-world deployment scenarios where high-powered machines might not always be available. Secondly, we leveraged the power of the DRDA-Net as the core network. DRDA-Net builds upon the strengths of CNNs by effectively extracting and utilizing features from the data. This combination empowers our model to achieve exceptional classification accuracy. By strategically combining a lightweight pre-trained model with a robust feature-learning network, we surpassed previous benchmarks on the BreaKHis dataset, paving the way for a more practical and accurate deep-learning solution for breast cancer diagnosis.